\documentclass[prd,twocolumn,preprintnumbers]{revtex4}

\usepackage{amsmath}
\usepackage{epsfig}
\usepackage{graphicx}
\usepackage{color}
\usepackage[normalem]{ulem}
 \usepackage{url}
\usepackage[breaklinks, plainpages=false, colorlinks=true, anchorcolor=cyan, linkcolor=red, citecolor=cyan, urlcolor=magenta, bookmarks=false]{hyperref}

\usepackage[caption=false]{subfig}

\begin{document}
\renewcommand{\thefigure}{\arabic{figure}}
\setcounter{figure}{0}

 \def\I{{\rm i}}
 \def\E{{\rm e}}
 \def\D{{\rm d}}

\bibliographystyle{apsrev}

\title{Rapid and Robust Parameter Inference for Binary Mergers}

\author{Neil J. Cornish}
\affiliation{eXtreme Gravity Institute, Department of Physics, Montana State University, Bozeman, Montana 59717, USA}

\begin{abstract} 

The detection rate for compact binary mergers has grown as the sensitivity of the global network of ground based gravitational wave detectors has improved, now reaching the stage where
robust automation of the analyses is essential. Automated low-latency algorithms have been developed that send out alerts when candidate signals are detected. The alerts include sky maps to facilitate electromagnetic follow up observations, along with probabilities that the system might contain a neutron star, and hence be more likely to generate an electromagnetic counterpart. Data quality issues, such as loud noise transients (glitches), can adversely affect the low-latency algorithms, causing false alarms and throwing off parameter estimation. Here a new analysis method is presented that is robust against glitches, and capable of producing fully Bayesian parameter inference, including sky maps and mass estimates, in a matter of minutes. Key elements of the method are wavelet-based de-noising, penalized maximization of the likelihood during the initial search, rapid sky localization using pre-computed inner products, and heterodyned likelihoods for full Bayesian inference.
\end{abstract}

\maketitle

\section{Introduction}

What started with a trickle in 2015~\cite{TheLIGOScientific:2016pea} has now turned into a veritable deluge~\cite{LIGOScientific:2018mvr,Abbott:2020niy} of gravitational wave signals detected by the LIGO and Virgo instruments. Keeping up with the ever increasing event rate is challenging. While the searches for gravitational wave signals are now highly automated and capable of producing near-real time alerts~\cite{Messick_2017,Klimenko_2016,canton2020realtime,Adams_2016,chu}, full parameter inference~\cite{Veitch_2015,Ashton_2019} has lagged behind. This is in part due to the large computational cost of fully Bayesian parameter inference, and in part due to the challenge of working with data that needs to be carefully calibrated~\cite{Viets_2018,virgocollaboration2018calibration} and cleaned of various noise contaminants~\cite{Driggers_2019,Davis_2019,Vajente_2020,Cornish:2020dwh}. 

The instrument noise in the LIGO and Virgo data is, for the most part, well described as locally stationary and Gaussian~\cite{LIGOScientific:2019hgc}. Short duration signals, such as high mass binary black holes, are in the sensitive band of the detector for a second or less, and the odds of the signal encountering a noise transient is low. However, for longer duration signals, such as low mass black hole binaries or systems that include a neutron star, the signals are in the sensitive band of the detectors for tens of seconds or even minutes, and it is likely that the signal will encounter a noise transient. As the low-frequency sensitivity of the detectors improve, low mass systems are virtually guaranteed to encounter noise transients. The LIGO/Virgo analyses have been fortified against noise transients using a combination of vetoes~\cite{Abbott_2018}, glitch robust search statistics~\cite{Allen:2004gu}, gating~\cite{Usman:2015kfa} and glitch subtraction~\cite{Cornish:2020dwh}. Gating of glitches is used in the online searches, while glitch subtraction is performed prior to off-line parameter estimation.

Here a proof of concept for a robust end-to-end low-latency Bayesian parameter estimation algorithm - {\tt QuickCBC} - is presented. The {\tt QuickCBC} algorithm reads in calibrated strain data, performs robust on-source spectral estimation, executes a rapid search for compact binary coalescence (CBC) signals,  uses wavelet de-noising to subtract any glitches from the search residuals, produces low-latency sky maps and initial parameter estimates, followed by full Bayesian parameter estimation.  For binary black holes the {\em entire} process takes just minutes on a laptop; for binary neutron stars, initial sky maps and mass estimates are ready in minutes, and full results are ready in less than an hour. The {\tt QuickCBC} code is open source~ \url{https://github.com/eXtremeGravityInstitute/QuickCBC}, and can be used to analyze public LIGO-Virgo data hosted by the Gravitational Wave Open Science Center \url{https://www.gw-openscience.org}. For testing purposes the algorithm was run in real time during the LIGO-Virgo O3 observing run, automatically generating results for triggers sent to the Gravitational-Wave Candidate Event Database. 

Existing algorithms can perform most of the individual steps in the {\tt QuickCBC} algorithm. For example, the {\tt PyCBC} search algorithm~\cite{Usman:2015kfa} performs low-latency searches for CBC signals that incorporate glitch mitigation via a chi-squared test~\cite{Allen:2004gu} and automatic gating of loud glitches~\cite{Usman:2015kfa}. The {\tt BayesWave} algorithm~\cite{Cornish:2014kda,PhysRevD.91.084034,Cornish:2020dwh}, produces on-source spectral estimates and performs glitch subtraction~\cite{TheLIGOScientific:2017qsa,Pankow:2018qpo,Abbott:2020niy}. The {\tt BayesStar} algorithm~\cite{Singer:2015ema} produces low-latency sky maps to help guide the search for electromagnetic counterparts to binary mergers, and several rapid parameter estimation algorithms have been developed~\cite{Pankow:2015cra,George:2017pmj,Wysocki:2019grj,Smith:2019ucc,Delaunoy:2020zcu}. What is novel is that {\tt QuickCBC} is a fully automated, end-to-end analysis algorithm that is robust against glitches, and able to produce reliable results in a matter of minutes. Many of the methods used by {\tt QuickCBC}, such as wavelet de-noising~\cite{1998BAMS...79...61T}, banded likelihoods for glitch rejection, and heterodyned likelihoods for rapid inference~\cite{Cornish:2010kf,Cornish:2020vtw}, are new to LIGO-Virgo data analysis.

Core elements of the {\tt QuickCBC} algorithm have been merged with the {\tt BayesWave} algorithm~\cite{BWCBC}. The key difference between the implementations is that the {\tt BayesWave} variant~\cite{Chatziioannou:2021ezd} jointly marginalizes over the CBC signal parameters, a model for the power spectral density (PSD) and a wavelet based model for noise transients. The {\tt QuickCBC} algorithm uses a fixed PSD and a point estimate for any noise transients. The other key difference is speed: {\tt QuickCBC} can be provide results with a latency of minutes, while the more refined {\tt BayesWave+CBC} analysis~\cite{Chatziioannou:2021ezd} takes hours.

\section{Overview of the {\tt QuickCBC} algorithm}

The {\tt QuickCBC} algorithm works with short snippets of LIGO-Virgo data, typically 4 to 8 seconds in length when searching for binary black holes and 16 to 32 seconds in length when searching for binary neutron stars. The run time scales roughly linearly with the data volume. 

The first step is to produce estimates for the power spectral density in each detector. On-source spectral estimation, where the short segment of data to be searched is also used to estimate the power spectral density (PSD), can be thrown off by the presence of loud signals or loud glitches. To avoid such biases, {\tt QuickCBC} uses an iterative approach that combines a running median estimate for the spectrum with spectral line identification and wavelet de-noising~\cite{1998BAMS...79...61T}. The de-noising removes signals and glitches, so only the spectral estimate from the first stage of the analysis is passes to the second stage. The second stage performs a rapid, network coherent search for CBC signals using a parallel tempered Markov Chain Monte Carlo algorithm (PTMCMC)~\cite{PhysRevLett.57.2607} with a banded likelihood that is analytically maximized over amplitude, phase and arrival time. Only the intrinsic parameters of the signal - masses and spins - are explored by the PTMCMC. The banded likelihood automatically identifies and rejects frequency bands that are impacted by noise transients. The removal is done separately for each time delay, resulting in a robust time-frequency glitch rejection method that can detect signals in the presence of glitches. The third stage subtracts the best-fit CBC waveform from the data, and performs a second round of spectral estimation and wavelet de-noising. The de-noising produces a glitch model that is subtracted from the original data, while preserving any gravitational wave signals. The cleaned data and updated spectral estimates are used in the subsequent stages of the analysis. The fourth stage refines the estimates of the intrinsic parameters using a standard non-maximized and non-banded likelihood function. Consequently, the amplitude, phase and arrival time at each detector also have to be explored by the PTMCMC. The refined estimates for the intrinsic parameters are then passed to the fifth stage of the analysis, which uses a PTMCMC algorithm with an algebraic likelihood function to map out the extrinsic parameters of the source - sky location, luminosity distance, inclination and polarization angles while holding the intrinsic parameters fixed. Since the extrinsic parameters only impact the projection of the waveform onto the detectors, the inner products in the likelihood can be pre-computed, resulting in an algebraic likelihood function that can be evaluated in a fraction of a microsecond~\cite{Cornish:2016pox}.  With the first five stages complete the algorithm will have produced a full three-dimensional sky map (RA, DEC and luminosity distance), along with estimates for the component masses and spins. All in about the time is took you to read this paragraph. The precise run time will depend on the duration and bandwidth of the signal and the speed and number of computations cores. For binary black hole systems at current LIGO/Virgo sensitivity  it is usually sufficient to use 4 seconds of data sampled at 2048 Hz. For neutron star - black hole systems it is enough to use 8 to 16 seconds of data sampled at 2048 Hz, while for binary neutron star systems we need 16 to 32 seconds of data sampled at 4096 Hz. For a binary black hole system, running on a 2016 MacBook Pro laptop with a 2.9 GHz quad-core processor, it takes $\sim 60$ seconds for the PSD estimation and intrinsic parameter search, and an additional $\sim 30$ seconds to complete the extrinsic parameter search and produce sky maps. The cost of the intrinsic parameter search scales linearly with the data duration, while the cost of the extrinsic parameter search scales linearly with the sample rate. Thus, the intrinsic parameter search for neutron-star black hole binaries takes either two to four times longer than for binary black holes, but time to produce a sky map is the same. Sky maps can be produced in very low latency by skipping the intrinsic search and instead using the intrinsic parameters provided by the search pipelines (as is done by  {\tt BayesStar}~\cite{Singer:2015ema}). The run time to produce a sky map with {\tt QuickCBC} is comparable to, or a little faster than, {\tt BayesStar}. The main difference is that the {\tt QuickCBC} maps are fully Bayesian, while {\tt BayesStar} maps are only approximately so.

\begin{figure}[htp]
\includegraphics[width=0.48\textwidth]{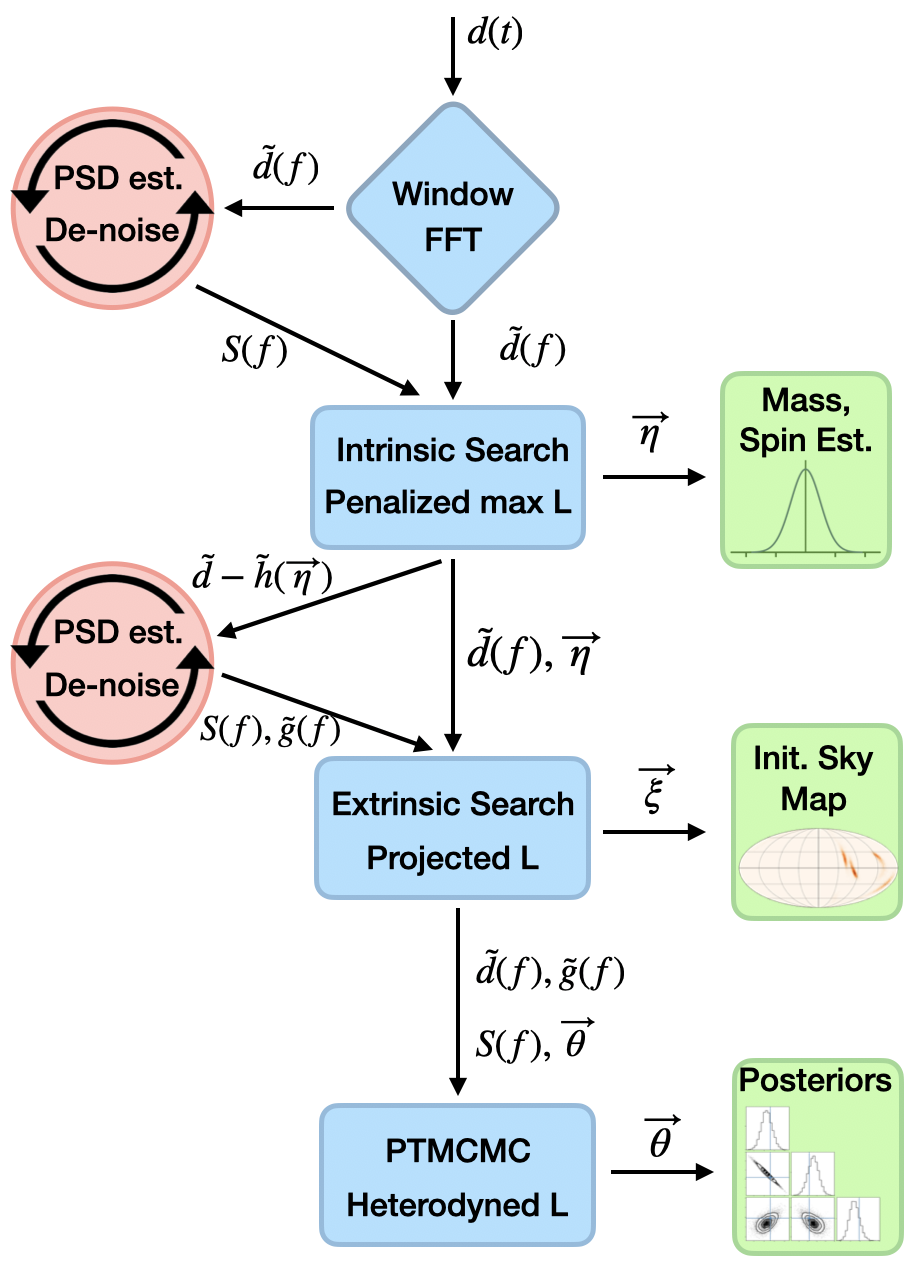} 
\caption{\label{fig:work} Workflow diagram for the {\tt QuickCBC} algorithm. The time domain data, $d(t)$, is read in, windowed, and Fourier transform to produce $\tilde d(f)$. An initial on-source PSD estimate $S(f)$ is produced using wavelet de-noising to remove glitch and signal power. Next the data is searched using a glitch-robust likelihood function that maximizes over extrinsic parameters and returns initial estimates for the intrinsic parameters $\vec{\eta}$. The PSD estimation is then repeated on the residual $d-h(\vec{\eta})$, and wavelet de-noising is used to fit and remove any glitches $\tilde{g}(f)$ that might be present in the data. The de-glitched data, $\tilde d(f) - \tilde g(f)$, is used to pre-compute various inner products for the projected network likelihood, allowing for a rapid mapping of the extrinsic parameters $\vec{\xi}$, such as the sky position and luminosity distance. The initial estimates for the full set of parameters $\vec{\theta} = \{\vec{\eta},\vec{\xi}\}$, are used along with the de-glitched data and refined PSD estimate to initialize the heterodyned likelihood function that is used by the PTMCMC algorithm to rapidly produce full posterior distribution.}
\end{figure}

The final stage of the algorithm refines the initial parameter estimates using a PTMCMC algorithm and a fast, heterodyned likelihood function~\cite{Cornish:2010kf,Cornish:2020vtw}. The heterodyned likelihood offers significant speed advantages, especially for long duration signals such as binary neutron star inspirals. The run-time for the final PTMCMC stage scales linearly with the sample rate and is independent of the observation time. On the same laptop computer described earlier, the PTMCMC stage takes four minutes for black hole binaries and neutron star - black hole binaries, and eight minutes for neutron star binaries. These run-times are several orders magnitude faster than the hours or days it takes for {\tt LALinference}~\cite{Veitch_2015,Ashton_2019} to produce results.

\subsection{Spectral Estimation and Wavelet De-noising}

The {\tt QuickCBC} algorithm is design to work with short stretches of data, typically between $T_{\rm obs} = 4$ and $T_{\rm obs} = 32$ seconds in duration. Traditional spectral estimation techniques, such as Welch averaging, can not be used on short data segments like these. The advantages of working with short data segments are speed and robustness against non-stationary drifts in the power spectrum. The disadvantages are low spectral resolution and possible biases due to the presence of loud signals or glitches. To guard against such biases an iterative wavelet de-noising approach is used to remove non-Gaussian features from the data.

\begin{figure}[htp]
\includegraphics[width=0.48\textwidth]{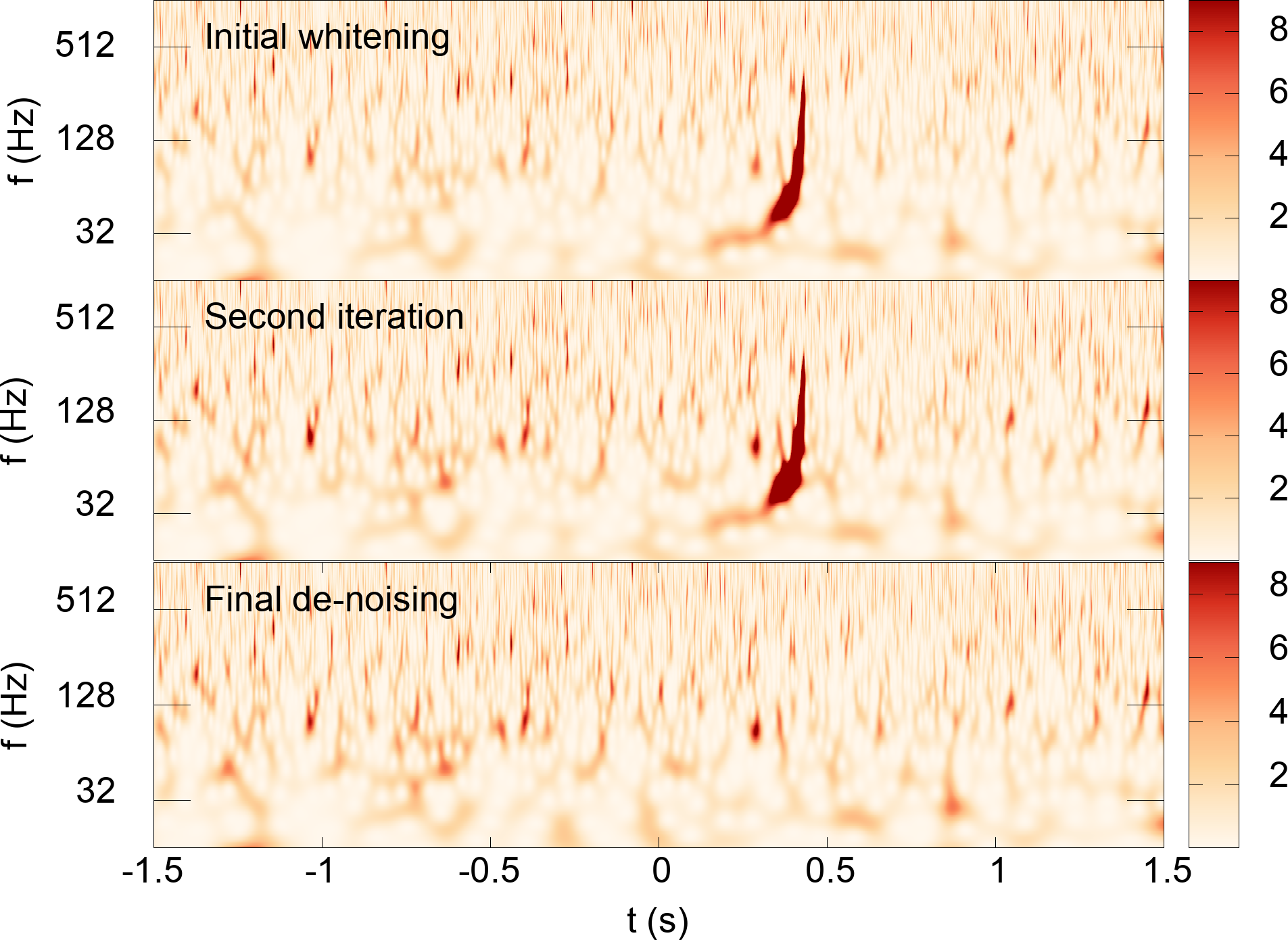} 
\caption{\label{fig:spec_est_GW150914} Time-frequency maps illustrating the spectral estimation and de-noising procedure applied to four seconds of LIGO Hanford data surrounding GPS time 1126259462. The initial spectral estimation and whitening (upper panel) is impacted by the loud signal from GW150914. Wavelet de-noising is used to remove the excess power, then the spectral estimation is repeated. The original data is re-whitened (middle panel) ready for the next iteration. The process is iterated until the excess SNR plateaus.  The bottom panel shows the de-noised data used to produce the final spectral estimate.}
\end{figure}

The iterative spectral estimation procedure proceeds as follows: A Tukey window is applied to the data to limit spectral leakage. A FFT is then used to produce a periodogram $S_p(f)$. A running median of width $\Delta f$ is used to smooth the periodogram. The width of the smoothing window $\Delta f$ is chosen to strike a balance between following the slope of the spectrum (small $\Delta f$'s follow the slope), and ignoring sharp spectral lines (large $\Delta f$'s are robust against lines). More accurately, it is the number of Fourier samples in the smoothing window that is critical, so for longer observation times smaller windows can be used. For the shortest 4 second segments the default width is $\Delta f = 16$ Hz. The smoothed spectral estimate $S_s(f)$ is used to identify spectral lines, which are found by taking the ratio $R(f)=S_p(f)/S_s(f)$, with lines defined as regions where this ratio exceeds $R_* =10$. The full spectral estimate is then given by
\begin{equation}
S(f) = \left\{\begin{array}{lr}
        S_s(f), & \text{for } R(f) \leq R_*\\
        S_p(f), & \text{for }  R(f) > R_* 
           \end{array}\right. \, .
\end{equation}
The initial spectral estimate is used to whiten the data: $\tilde{d}(f) \rightarrow \tilde{d}_w(f) = \tilde{d}(f)/\sqrt{S(f)}$.  The whitened data is wavelet transformed using an over-complete collection of Morlet-Gabor continuous wavelets (see the top panel of Figure~\ref{fig:spec_est_GW150914}). A wavelet de-noising procedure~\cite{1998BAMS...79...61T} is then used to remove any non-Gaussian features from the data. Wavelet de-noising is basically a time-frequency thresholding technique. For stationary Gaussian noise, the wavelet power spectrum, $S_{nm}$, should follow a chi-squared distribution with two degrees of freedom. Wavelet pixels with power above a certain threshold are identified, then an inverse wavelet transform of these pixels is used to produce a whitened time domain reconstruction of the excess power. The reconstructed feature is re-colored using the smooth component of the power spectrum $S_s(f)$ and subtracted from the original time domain data. The thresholding procedure starts by identifying pixels with $S_{nm} > S_0$, then surrounding pixels with $S_{n\pm1 m\pm1} > S_1$ are also flagged, with the goal of identifying clusters of excess power. The standard threshold values are $S_0=10$ and $S_1=6$. The power spectral estimation is repeated using the de-noised data. The updated power spectrum is then used to whiten the original data, and the wavelet de-noising procedure is repeated. The entire procedure is iterated until the signal-to-noise of the non-Gaussian excess plateaus. This typically takes just one or two iterations, but data with very loud glitches may require as many as five or six iterations for the procedure to converge.

\begin{figure}[htp]
\includegraphics[width=0.48\textwidth]{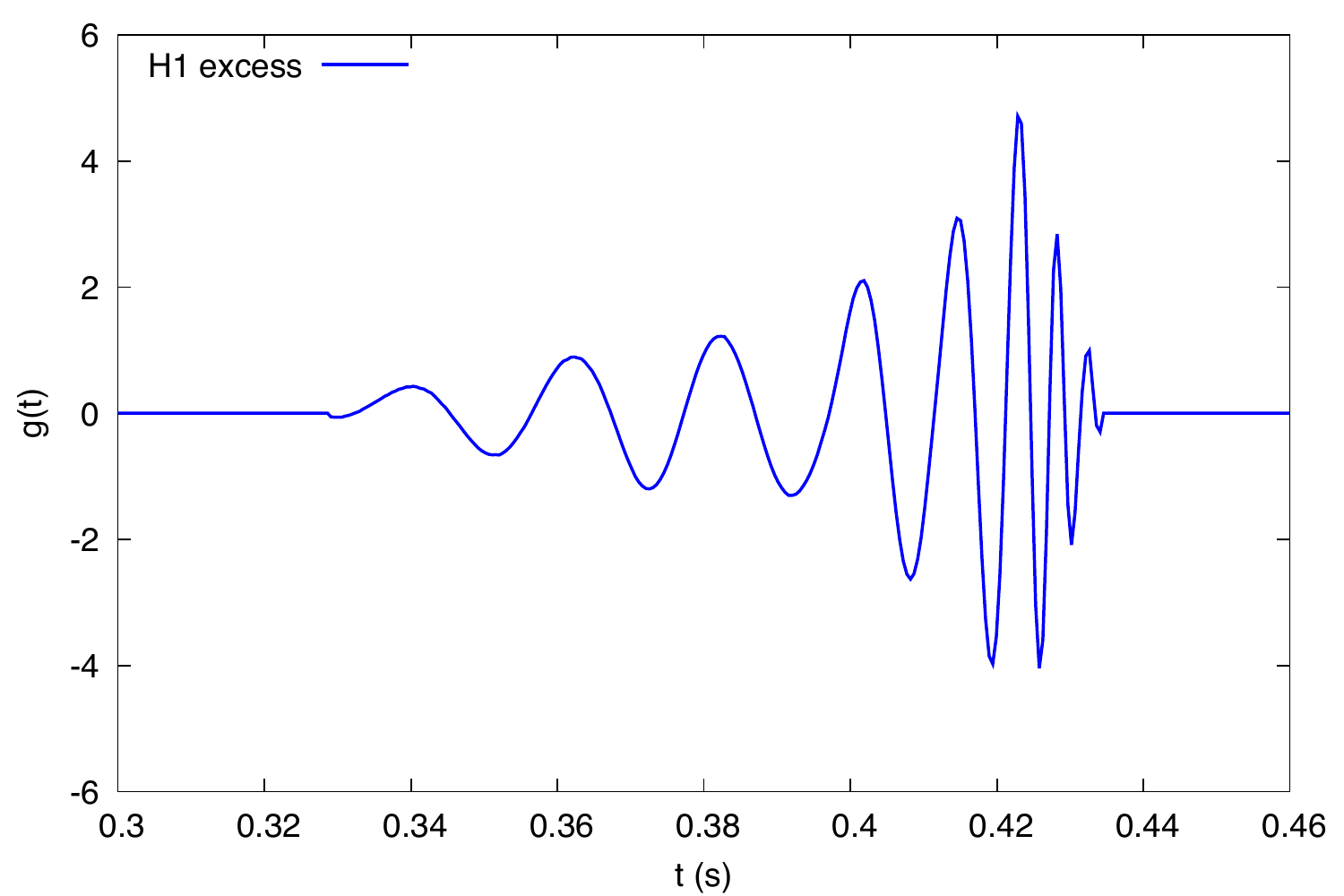} 
\caption{\label{fig:excess_GW150914} The non-Gaussian feature removed by wavelet de-noising during the iterative spectral estimation procedure applied to four seconds of LIGO Hanford data centered on GPS time 1126259462. In this instance the non-Gaussian feature is the gravitational wave signal from the binary black hole merger GW150914.}
\end{figure}

Figure~\ref{fig:spec_est_GW150914} illustrates the spectral estimation and de-noising procedure using 4 seconds of data surrounding GW150914. The initial spectral estimate used to whiten the data (top panel) overestimates the height of the spectrum through the band between $\sim 40 {\rm Hz} \rightarrow 120 {\rm Hz}$ due to the loud gravitational wave signal. The data in the middle panel is whitened using the updated spectrum found after the initial round of wavelet de-noising. More features are now visible in the mid-band frequencies. In this example the procedure converged after two iterations. The lower panel shows the de-noised data that was used to produce the final spectral estimate. Figure ~\ref{fig:excess_GW150914} shows the whitened time domain reconstruction of the feature that was removed by the de-noising process. In this instance the feature is the gravitational wave signal from a binary black hole merger. It would be a bad idea to use the wavelet de-noised data from the spectral estimation procedure for subsequent stages in the analysis! Instead, just the spectral estimate is used.

\begin{figure}[htp]
\includegraphics[width=0.48\textwidth]{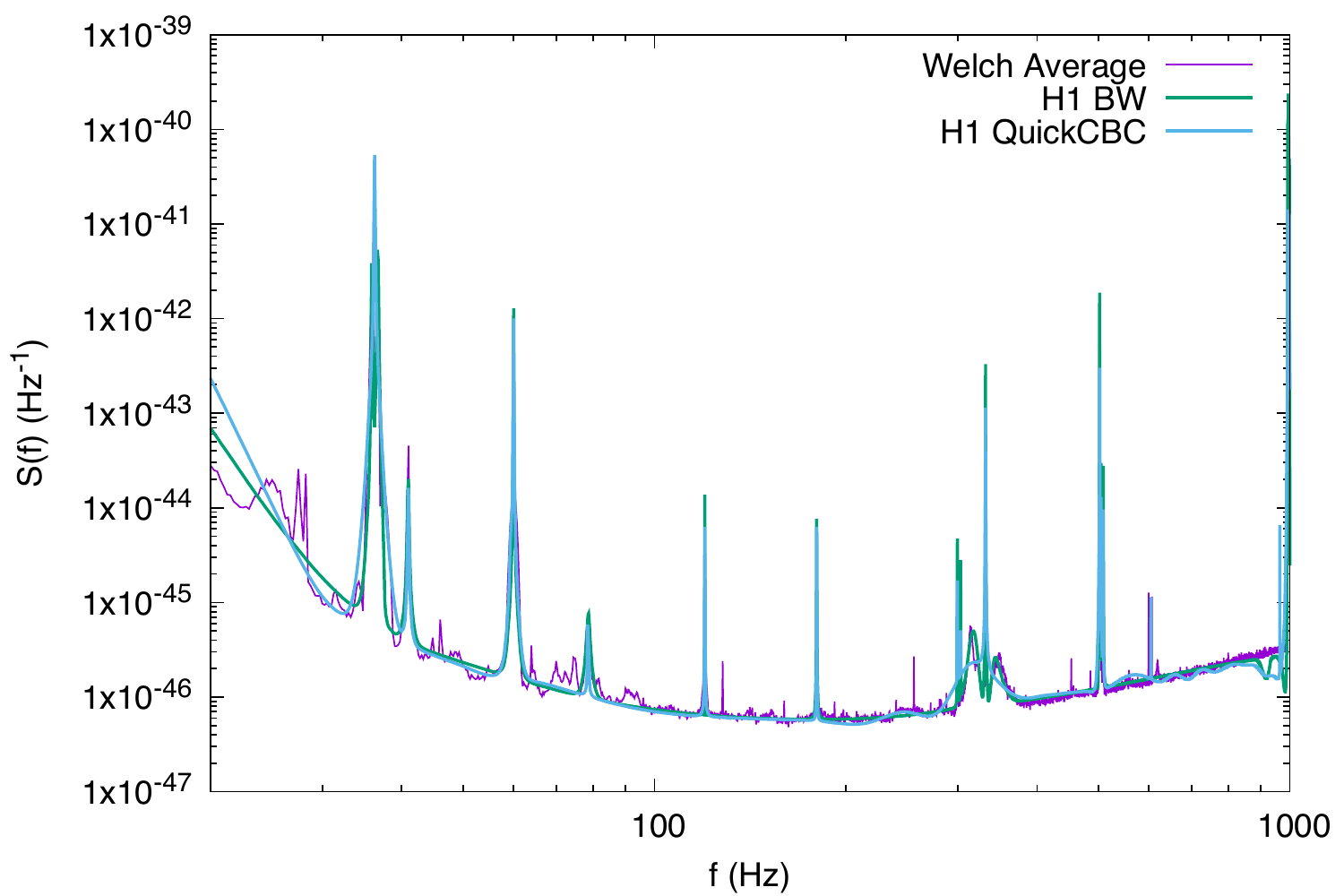} 
\caption{\label{fig:PSD_GW150914} A comparison of power spectral density estimates using four seconds of LIGO Hanford data surrounding GW150914 centered on GPS time 1126259462. The low-latency {\tt QuickCBC} algorithm produces a good approximation to the reference {\tt BayesWave} estimate. A Welch average using 2048 seconds of data surrounding the event is shown for comparison.}
\end{figure}

The final stage of the {\tt QuickCBC} spectral estimation procedure is to fit a fixed dimension version of the {\tt BayesWave} trans-dimensional spectral model~\cite{Cornish:2014kda,PhysRevD.91.084034} to the glitch-subtracted data. The spectral model includes a smooth component described by a cubic spline, and line features described by a Lorentzian line model. The running median is used to initialize the spline model. The spacing of the spline points is determined by comparing two running averages of the running median, one with a window twice as wide as the other. The usual choice is to use 4 Hz and an 8 Hz window. The spline control points are spaced more closely in regions where the two averages diverge, and spaced further apart in regions where the two averages converge. The minimum spacing of the spline control points is set at 4 Hz and the maximum spacing is set at 32 Hz. The threshold on the ratio of the the two averages is set at 20\%. The Lorentzian line model is initialized using the outliers from the running median to set the initial location, amplitude and width of the lines. A simple fixed dimension MCMC is then used to refine the model parameters. To control unphysical oscillations in the spline model, a prior is used that penalizes points with large second derivatives. Figure~\ref{fig:PSD_GW150914} compares the {\tt QuickCBC} spectral estimate to the estimate from the {\tt BayesWave} algorithm~\cite{Cornish:2014kda,PhysRevD.91.084034}. The {\tt QuickCBC} estimate agrees very well with the {\tt BayesWave} estimate. The key difference is that the {\tt QuickCBC} estimate is produced in seconds, while the more refined {\tt BayesWave} estimate takes tens of minutes or longer to produce.

\subsection{Parallel Tempered Markov Chain Monte Carlo}

The {\em QuickCBC} algorithm uses a Parallel Tempered Markov Chain Monte Carlo (PTMCMC) algorithm~\cite{PhysRevLett.57.2607,Littenberg_2009,Vousden_2015} for both the initial search, the fast sky mapping and the full parameter inference. The implementation varies slightly between the different stages, mostly in terms of the likelihood function used, but the central engine is the same. 

The PTMCMC algorithm uses a collection of chains that explore likelihoods that are scaled by ``inverse temperatures'' $\beta_i$: $\ln L(\beta_i, \vec{\theta}) = \beta_i \ln L (\vec{\theta})$. Chains with $\beta_i=1$ explore the target posterior, while chains with $\beta_i > 1$ range more widely, and help chains to escape from local maxima. Chains with $\beta_i < 1$ can be used to help lock on to weak signals. The temperature ladder is set so that there are multiple ``cold'' chains: $\beta_i = 1$ for $i=[1,N_{\rm c}]$ that interact with $N_h$ ``hot'' chains on a geometrically spaced temperature ladder: $\beta_k = \alpha^{k}$ for $k=[1,N_{\rm h}]$. The choice of increment $\alpha$ and the number of hot chains are tunable parameters. Setting $\alpha$ too large will result in poor exchange between chains. Setting $\alpha$ too small will require a prohibitive number of chains to reach the desired maximum temperature. A good rule-of-thumb for the maximum temperature is that the effective signal-to-noise ratio for the hottest chain, ${\rm SNR}_{N_h} = {\rm SNR}\sqrt{\beta_{N_k}}$ should be of order 4-5. For typical LIGO-Virgo signals with ${\rm SNR} \sim 10 \rightarrow 20$ we need $\beta_{N_k} \sim 1/4 \rightarrow 1/16$. Using a spacing of $\alpha = 0.8$ requires of order $N_h \sim 6 \rightarrow 12$ hot chains to reach the desired maximum temperature. These rule-of-thumb settings for the PTMCMC temperature ladder have been found to work in practice, but the efficiency could be improved by using dynamic temperature spacing~\cite{Vousden_2015}. When loud signals (${\rm SNR} > 30$) are detected, it may be necessary to start a new analysis with additional chains to have a high enough maximum temperature.

Each chain is updated using a mixture of proposal distributions. The standard mix includes draws from the prior distribution, jumps along eigenvectors of the Fisher information matrix, differential evolution, Gaussian jumps along each intrinsic parameter direction and a dedicated extrinsic parameter proposal that draws new sky locations that maintain the time delay between a randomly selected pair of detectors, while analytically adjusting the other extrinsic parameters to keep the detector frame waveforms unchanged. The extrinsic proposal is described in detail in section IVa of Ref.~\cite{Cornish:2020dwh}. 

The Fisher information matrix proposal is based on a quadratic expansion of the log likelihood:
\begin{eqnarray}
&& \Gamma_{ij}(\vec{\theta}) = -\partial_i \partial_j \ln L \nonumber \\
&& \quad  = 4 \sum_{ab} \int \frac{{\cal A}^{ab}_{,i}{\cal A}^{ab}_{,j}+{\cal A}_{ab}^2 \Phi^{ab}_{,i}\Phi^{ab}_{,j}}{S^{a}(f)} \, df \, .
\end{eqnarray}
Here the derivatives are taken with respect to the waveform parameters $\theta^i$ centered on some reference value $\vec{\theta}$. The sum is over the detectors, $a$, in the network, and the harmonics, $b$, of the gravitational wave signal:
\begin{equation}
h_a(f) = \sum_b  {\cal A}_{ab}(f) e^{i \Phi^{ab}(f)} \, .
\end{equation}
The reference value $\vec{\theta}$ is updated to the current value of the chain every few hundred iterations and the Fisher matrix is recomputed at the new location. The Fisher matrix proposal employs the eigenvectors ${\bf v}_{(k)}$ and eigenvalues $\lambda_k$, found by solving the linear system
\begin{equation}
 \Gamma_{ij} v^j_{(k)} = \lambda_k v^i_{(k)} \, .
\end{equation}
Jumps from the current location ${\bf x}$ to candidate location ${\bf y}$ are proposed by first randomly selecting an eigen-direction $p$, and setting
\begin{equation}
{\bf y} = {\bf x} + \frac{\gamma}{\sqrt{\lambda_p}} {\bf v}_{(p)} \, ,
\end{equation}
where $\gamma \sim {\cal N}(0,1)$ is a zero mean, unit variance Gaussian deviate. The proposal densities for this jump cancel in the Metropolis-Hastings ratio since the Fisher matrix is held fixed (aside from occasional infrequent updates). Jumping along eigen-directions is more robust that drawing from the full Fisher matrix, as the matrices are often poorly-conditioned. The poor conditioning typically only impacts one or two eigen-directions,  and still allows for good acceptance of jumps along the other eigen-directions. Small Gaussian jumps along individual parameter directions are included in the proposal mix to help cover directions that might not be explored well by the Fisher matrix jumps.

Differential evolution (DE) proposals~\cite{ter2006markov} are particularly good at exploring degenerate directions in parameter space - the very same directions that cause the Fisher matrix to become ill-conditioned. The variant of differential evolution used by {\tt QuickCBC} works as follows: A history array of past samples, ${\bf z}$ is collected for each temperature level (with multiple copies for the cold chains). The array is initialized with draws from the prior. Samples are added to the history array after every $\sim 10$ iterations. A counter $j$ keeps track of how many samples have been added. The new sample added is added to the array at index $j({\rm mod}N_H)$. That is, when $j$ reaches $N_H$ the first entry in the array gets replaced and so on. The DE proposal is made as follows: Two samples, $k,l$, are drawn from the history array and used to propose a new location
\begin{equation}
{\bf y} = \bf{x} + \gamma(\bf{z}_k-\bf{z}_l) \, .
\end{equation}
Here $\gamma$ is drawn from a Gaussian of width $2.38/\sqrt{2 d}$ for 90\% of the DE updates, where $d$ is number of parameters, and set to $\gamma =1$ for the rest. The proposal is symmetric, so the proposal densities cancel in the Metropolis-Hastings ratio. The Gaussian DE jumps are good for exploring local correlations, while the $\gamma =1$ DE jumps allow the chains to move between discrete modes of the posterior.

The {\tt QuickCBC} sampler is currently limited to using waveform templates that describe non-precessing, quasi-circular binaries. These templates can be parameterized in terms of four intrinsic parameters and seven extrinsic parameters. The intrinsic parameters are the individual mass $m_1,m_2$ and the aligned dimensionless spins $\chi_1, \chi_2$. Here $\chi = \vec{\chi} \cdot \hat{L}$, where $ \vec{\chi} = \vec{S}/m^2$ is the dimensionless spin vector and $\hat{L}$ is a unit vector aligned with the orbital angular momentum. The seven extrinsic parameters are the sky location RA, DEC = $(\alpha,\beta)$, luminosity distance $D_L$, polarization and inclination $(\psi,\iota)$, merger time and merger phase $(t_c,\phi_c)$. 

The  {\tt QuickCBC}  sampler uses the modified collection of parameters $\vec{\eta} \rightarrow \{\ln {\cal M}, \ln M, \chi_1, \chi_2\}$ and $\vec{\xi} \rightarrow \{\alpha, \sin \beta, \ln D_L, \psi, \cos \iota, \phi_c, t_c,\}$, where $M = m_1+m_2$ is the total mass and ${\cal M} = (m_1 m_2)^{3/5}/M^{1/5}$ is the chirp mass. The priors are taken to be uniform in all the parameters save for ${\cal M}, M, D_L$. For the masses the priors are uniform in $m_1,m_2$, which can be enforced using the Jacobian factor $J_M = M m_1 m_2  /\sqrt{M^2 - 4 m_1 m_2}$. For the distance, the prior is taken to be uniform in luminosity distance volume, which can be enforced using the Jacobian factor $J_D = D_L^3$. Some waveform models, such as the IMRPhenomD model~\cite{Husa:2015iqa,Khan:2015jqa} used to produce the plots in this paper, are only considered to be reliable for a sub-set of mass ratios and spins. To account for this, the prior ranges for the IMRPhenomD analyses are restricted such that $m_1/m_2 < 18$ and $| \chi | <  \chi_{\rm max} = 0.85$. The default prior on the spins is uniform in the aligned spin component. To facilitate comparison with the IMRPhenomPv2 precessing model~\cite{Hannam_2014}, which uses a uniform-in-direction spin prior, a second spin prior option can be selected that is uniform in the aligned spin component for isotopically distributed spins, namely, $p(\chi) = \ln( \chi_{\rm max} / |\chi|)/(2 \chi_{\rm max})$.

\subsection{Glitch Robust Coherent Search}

The {\tt QuickCBC} algorithm can be used to search for CBC signals in segments of LIGO/Virgo data. The standard usage is to follow-up triggers from template-bank based CBC search pipelines, but any valid GPS time will do. {\tt QuickCBC} executes a stochastic search using a PTMCMC algorithm and a glitch-robust maximized likelihood function. The search is limited to the dominant waveform harmonic for non-precessing, quasi-circular binaries. As such, the search may fail to detect systems with significant contributions from higher modes, strongly precessing systems, or highly eccentric systems. Extending the search to include higher modes is straightforward. Including precession and eccentricity is far more challenging.

The dominant waveform harmonic for non-precessing, quasi-circular binaries has polarization states related: $h_\times(f) = \I \epsilon h_+(f)$ where
\begin{equation}
\epsilon = -\frac{2 \cos\iota}{(1+\cos^2 \iota)} \, .
\end{equation}
The detector response can be written as:
\begin{equation}
h_a(\vec{\theta}, f) = h_+(\vec{\eta}, f) \frac{D_*}{D_L}\left(F^a_+  + \I \epsilon F^a_\times\right) \E^{2\pi \I f \Delta t_a} \E^{\I \phi_c}
\end{equation}
where $a$ labels the detector, $F^a_+(\alpha,\beta,\psi)$ and $F^a_\times(\alpha,\beta,\psi)$ are the antenna response patterns, $\Delta t_a$ is the arrival time relative to the geocenter time, $\phi_c$ is the merger phase, and $D_L$ is the luminosity distance. The reference geocenter waveform $h_+(\vec{\eta},f)$ is generated using an arbitrary fiducial luminosity distance $D_*$, with merger time and phase set equal to zero. As such, the reference waveform only depends on the four intrinsic parameters $\vec{\eta}$. Defining
\begin{equation}
F_a = \frac{D_*}{D_L} \left({F^a_+}^2 + \epsilon^2 {F^a_\times}^2\right)^{1/2}
\end{equation}
and
\begin{equation}\label{phase}
\lambda_a = {\rm atan}(\epsilon F^a_\times/F^a_+) + \phi_c
\end{equation}
the response can be written as
\begin{equation}\label{sig}
h_a(\vec{\theta}, f) = h_+(\vec{\eta}, f)  F_a \E^{\I \lambda_a} \E^{2\pi \I f \Delta t_a} \, .
\end{equation}
We see that the waveforms in each detector are identical up to an overall amplitude scaling, time shift and phase shift.  Denoting the data in detector $a$ as $d_a(f)$,
the Gaussian log likelihood is given by
\begin{equation}\label{logL}
\ln L_a = (d_a \vert h_a)  - \frac{1}{2} (h_a|h_a)  - \frac{1}{2} (d_a|d_a)\, .
\end{equation}
Here $(x\vert y)$ denotes the noise weighted inner product
\begin{equation}
(x\vert y) = 2 \int \frac{ x^* y + y^* x}{S(f)} \, d f \, ,
\end{equation}
where $S(f)$ is the power spectral density of the noise. The network log likelihood is found by summing the individual contributions: $\ln L = \sum_a \ln L_a$. 

\subsubsection{Maximized likelihood}

During the search stage, the power spectral density is held fixed and the $(d_a|d_a)$ term is a constant that can be ignored. Standard tricks are used to maximize over the amplitude, phase and arrival time of the waveforms. Writing the signal in terms of unit normalized sine and cosine quadratures, $(h_{a,s}|h_{a,s})=(h_{a,c}|h_{a,c})=1$, $(h_{a,s}|h_{a,c})=0$:
\begin{equation}
h_a(f) = A_a \left( h_{a,s}(f) \sin \phi_a+   h_{a,c}(f) \cos \phi_a  \right) \, ,
\end{equation}
the log likelihood (dropping the $(d_a|d_a)$ constant term) becomes
\begin{equation}\label{logLq}
\ln L_a = A_a \rho_a(t_a,\vec{\eta}) \cos(\phi_a - \varphi_a(t_a,\vec{\eta})) - \frac{1}{2} A_a^2 \, ,
\end{equation}
where $\vec{\eta}$ are the intrinsic parameters of the source, and $\rho_a(t_a,\vec{\eta}) = |z_a(t_a,\vec{\eta})|$, $\varphi_a(t_a,\vec{\eta}) = {\rm arg}\{ z_a(t_a,\vec{\eta})\} $ where
\begin{equation}\label{zs}
z_a(t_a, \vec{\eta})  = 4 \int \frac{ d_a(f) h_{a,c}^*(f, \vec{\eta}) }{S_n(f)} e^{2\pi \I f t_a} \, .
\end{equation}
The likelihood is maximized with respect to amplitude and phase by setting $A_a = \rho_a$ and $\phi_a = \varphi_a$:
\begin{equation}\label{logLmax}
\ln L_{a,{\rm max}}(t_a,\vec{\eta}) =  \frac{1}{2} \rho^2_a(t_a,\vec{\eta})  \, .
\end{equation}
The complex SNR time series $z_a(t_a, \vec{\eta})$ can be computed using an inverse fast Fourier transform. The likelihood can then be maximized with respect to the time offset $t_a$ by sorting the resulting time series $\rho_a(t_a,\vec{\eta})$. The network likelihood
\begin{equation}\label{logLnetmax}
\ln L_{\rm max}(\{ t_{i}\}, \vec{\eta})=  \frac{1}{2} \sum_a  \rho^2_a(t_a,\vec{\eta})
\end{equation}
can be maximized with respect to the arrival times in each detector, $\{ t_i \}$, subject to the constraint that the time differences $\Delta t_{ij} = | t_i - t_j|$ are less than the light travel times between the detector sites. The maximization is done pair-wise between detectors, starting with a reference detector. For networks with three or more detectors the pair-wise approach can yield collections of time delays that do not correspond to any physical sky location. Similarly, the relative phases may not correspond to any physical sky location, inclination and polarization angle. In most cases this is not a problem as the extrinsic parameters get refined in the subsequent sky-mapping stage of the analysis.

When glitches are present in the data, the log likelihood times series in each detector, $\ln L_{a,{\rm max}}(t_a,\vec{\eta})$, may have multiple distinct maxima. Some of these maxima will be associated with glitches and some will be associated with the signal. To account for this possibility, all maxima that are at least 50 ms apart are recored for each detector before applying the network time-delay restriction. The algorithm can return multiple solutions, each with different arrival times, amplitudes and phases. A glitch rejection step is then applied to each candidate solution before arriving at a unique maximum likelihood solution.

\subsubsection{Banded Glitch Rejection}

CBC searches use variants of the $\rho$ search statistic, defined for data $d$ and templates $h$ as
\begin{equation}
\rho = \frac{(d|h)}{(h|h)^{1/2}} \, .
\end{equation}
When a glitch is present in the data, $d=n+g$, the template can ring-off against the glitch. Usually this occurs across a narrow band of frequencies. Looking at how $\rho$ accumulates with frequency can be used to detect glitches. Rather than steadily accumulating, $\rho$ gets a big boost in the frequency band where the signal crosses a glitch. Motivated by these considerations, a chi-squared test for glitch rejection has been incorporated into CBC searches~\cite{Allen:2004gu}. The statistic uses frequency bands on varying width, with the width chosen so that the template has equal SNR=$\sqrt{(h|h)}$ in each band. 

\begin{figure}[htp]
\includegraphics[width=0.48\textwidth]{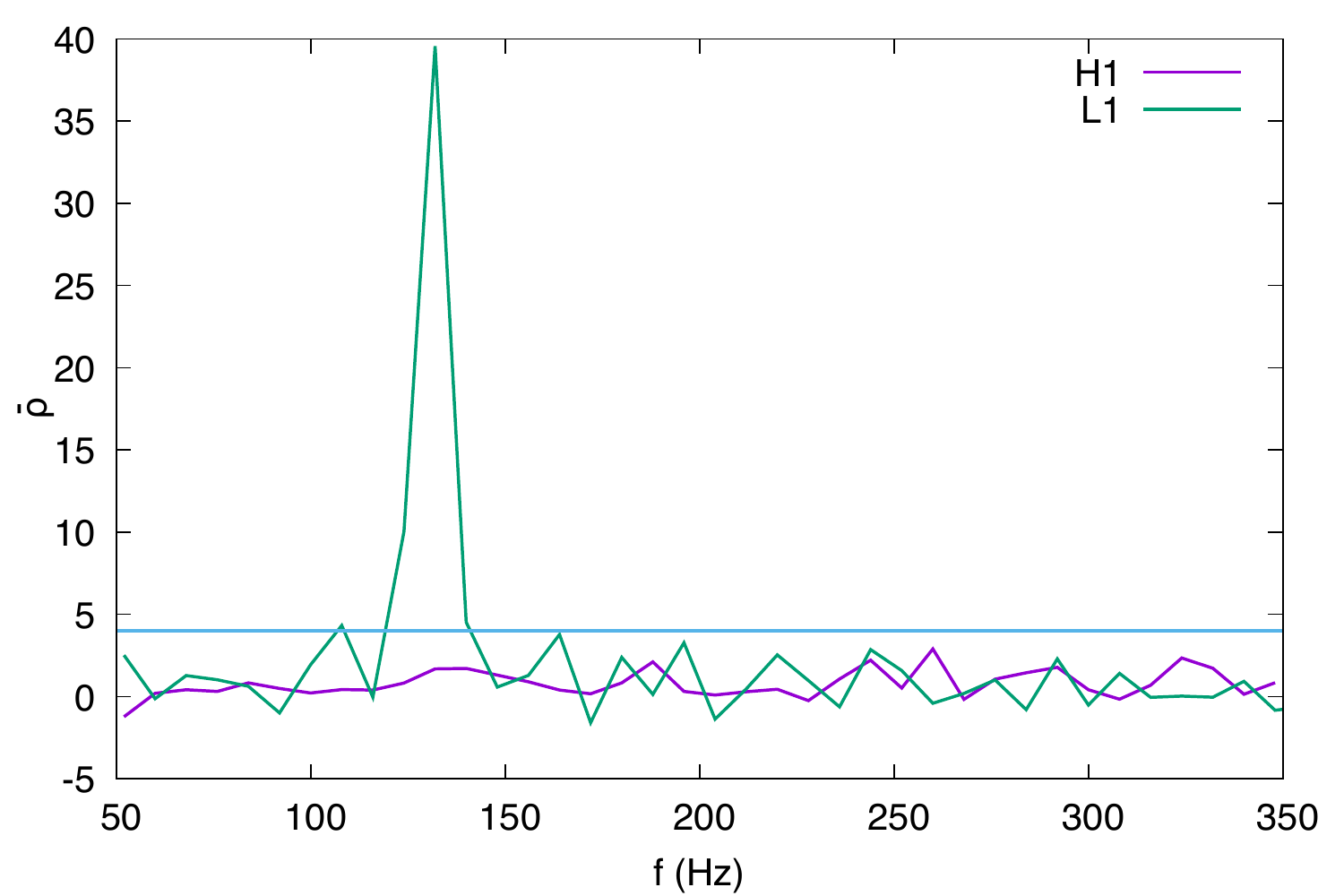} 
\caption{\label{fig:rhoGW170817} The banded $\bar{\rho}$ statistic as a function of central frequency for a template with chirp mass ${\cal M}=1.197 M_\odot$, total mass $M=2.8 M_\odot$ and merger time $t_c=1187008882.4486$ GPS seconds. The $\bar{\rho}$ statistic successfully identifies the frequency bands  where the signal track crosses a loud noise transient in the LIGO Livingston detector. }
\end{figure}

Here we introduce a variant of this approach with fixed-width frequency bands. Defining the ${\bar \rho}$ statistic:
\begin{equation}
{\bar \rho}(f,\Delta f) = \frac{(d-h|h)_{{\rm max} \phi_0}}{(h|h)^{1/2}}\, 
\end{equation}
where the noise-weighted inner products are computed across a frequency band of width $\Delta f$, centered at frequency $f$. The inner product of the residual, $d-h$, and the template, $h$, is analytically maximized with respect to the overall phase in that band using sine/cosine quadratures. In pure Gaussian noise, $d=n$, we have
\begin{equation}
{\rm E}[{\bar \rho}] = 1-{\rm SNR},\quad  {\rm Var}[{\bar \rho}]=1.
\end{equation}
When the template matches a signal in the data, $d=n+h$, we have
\begin{equation}
{\rm E}[{\bar \rho}] = 1,\quad  {\rm Var}[{\bar \rho}]=1.
\end{equation}
When a glitch is present in the data, $d=n+g$ the template rings-off against a glitch and ${\bar \rho}(f,\Delta f)$ becomes large and positive. Frequency bands where ${\bar \rho}(f,\Delta f) > 4$ are excluded from the likelihood calculation. The amplitude and phase maximization are repeated for the full template with any glitch-impacted bands removed.
The banded glitch rejection is applied to the collection of candidate solutions from the original likelihood maximization step. The solution with the largest banded likelihood is returned and used by the PTMCMC search algorithm.

\subsection{Glitch Removal}

\begin{figure}[htp]
\includegraphics[width=0.48\textwidth]{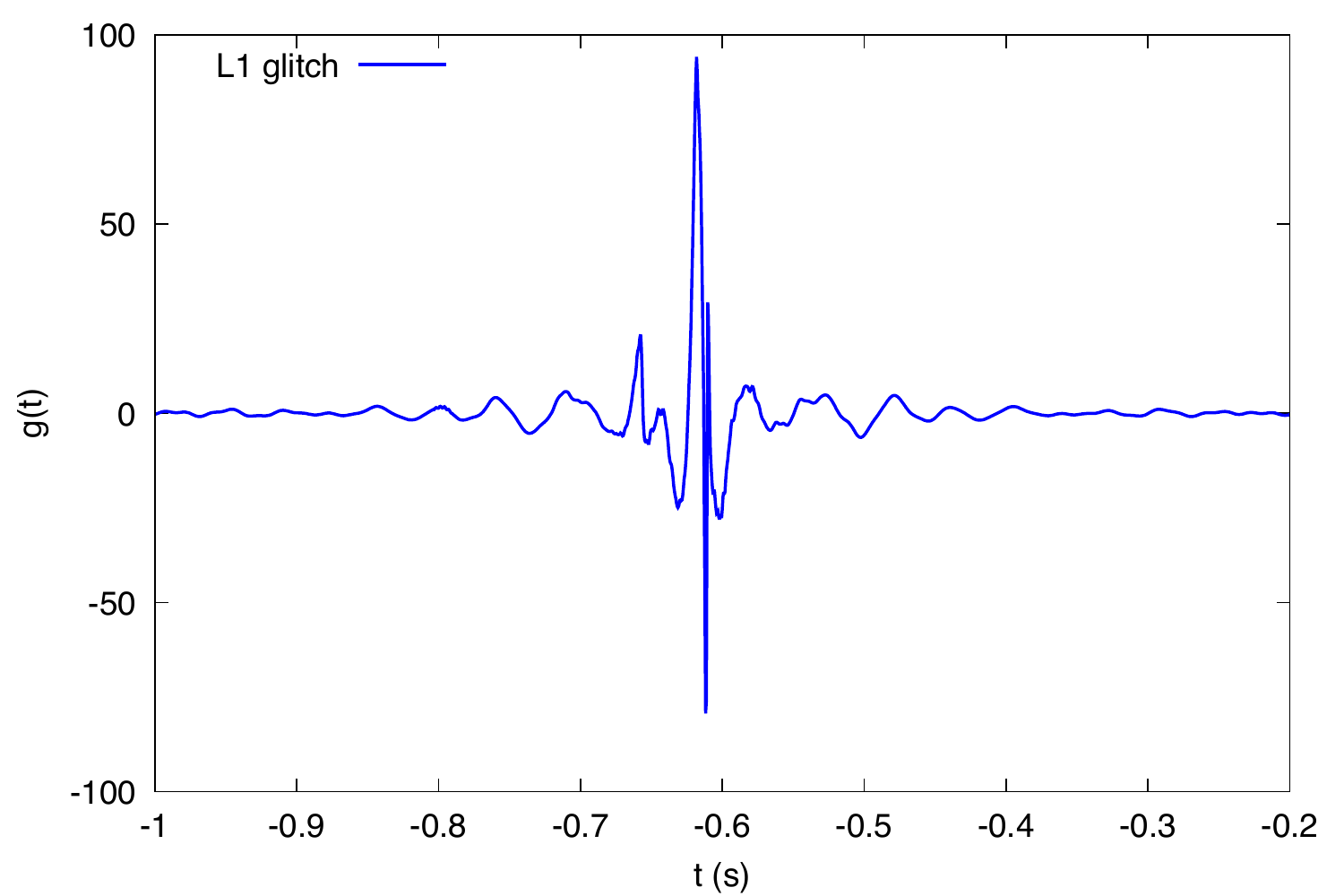} 
\caption{\label{fig:glitchGW170817} The whitened glitch model in LIGO Livingston data centered at GPS time 1187008882. The glitch reconstruction was performed after a low-latency point estimate for the GW170817 signal, which is coincident with the glitch, was subtracted from the data.}
\end{figure}

The PTMCMC search using the banded maximum likelihood function returns an initial estimate for the extrinsic parameters of the signal, along with the arrival times, amplitudes and phases in each detector. This solution is then used to subtract the CBC signal from the data. The residual is then processed through the same spectral estimation procedure that was applied to the original data. Figure~\ref{fig:glitchGW170817} shows the reconstructed glitch model for residual in the LIGO Livingston detector roughly a second before the merger of binary neutron star GW170817. 

\begin{figure}[htp]
\includegraphics[width=0.48\textwidth]{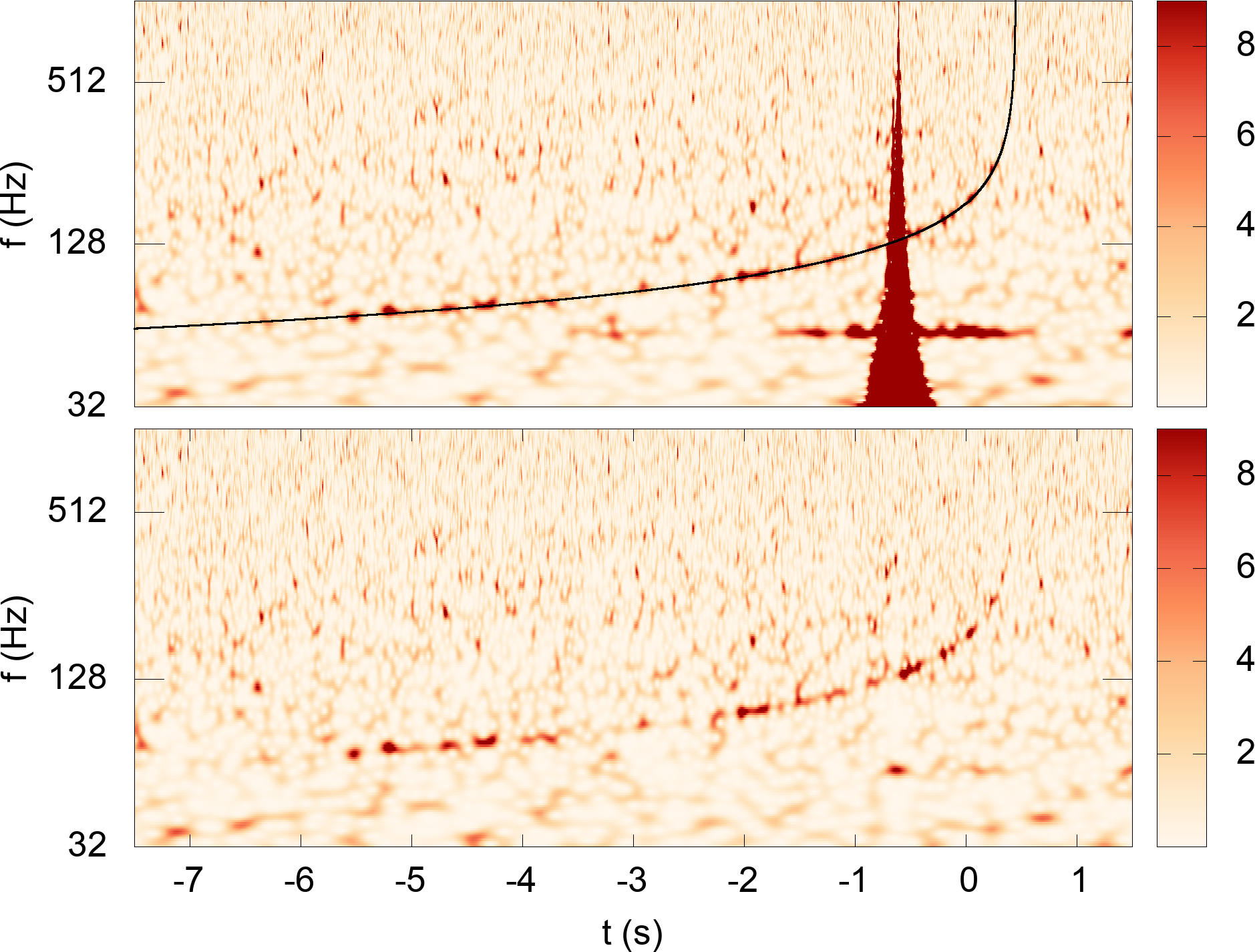} 
\caption{\label{fig:scanGW170817} Time-frequency maps of the LIGO Livingston data centered at GPS time 1187008882. The upper panel shows the raw whitened data. The black line indicates the reconstructed time-frequency track for GW170817 found using the banded maximum likelihood. The best-fit signal is subtracted from the data, then wavelet de-noising is used to identify any noise transients in the data. The noise transients are removed from the original data in preparation for more refined parameter estimation (lower panel).}
\end{figure}

Figure~\ref{fig:scanGW170817} shows time-frequency maps of the LIGO Livingston data surrounding the GW170817 event. The upper panel shows the time-frequency track for the point estimate of the signal that was subtracted from the data prior to the second round of spectral estimation and wavelet de-noising. The lower panel shows the whitened data after the noise transient has been removed. All subsequent stages of the analysis are performed using the glitch subtracted data.

\subsection{Low Latency Sky Mapping}

The search phase delivers an estimate for the intrinsic parameters, in addition to the amplitudes, phases and arrival times in each detector. The next step is to find extrinsic parameters that are consistent with the waveforms seen in each detector. With three or more detectors the problem of solving for the extrinsic parameters  $\vec{\xi}$ given the relative amplitudes, arrival times and phases is over-constrained, and often ill-posed due to noise. Rather than trying to solve the problem analytically, we once again resort to a Monte Carlo approach, this time aided by an extremely cheap-to-compute likelihood function.

When the intrinsic parameters $\vec{\eta}$ are held fixed, the response in each detector can be found by applying projections to a reference geocenter waveform that amount to amplitude re-scalings, phase shifts and time shifts (see equation~\ref{sig}). Taking a reference waveform $\hat{h}_+$, scaled to unity at some reference distance $D_a$, the log likelihood can be written as 
\begin{equation}\label{logLsky}
\ln L = \sum_a \frac{F_a D_a}{D_L} \left( e^{-\I \lambda_a} C(t_a) + e^{\I \lambda_a} C^*(t_a) \right)  - \frac{F_a^2 D_a^2}{2 D^2_L} \, .
\end{equation}
where
\begin{equation}\label{Ce}
C_a(t_a) =   \int \frac{ d_a \hat{h}_+^* }{ S_n(f)} \,  \E^{2\pi \I f t_a} d f  \, ,
\end{equation}
can be evaluated using a Fast Fourier transform (FFT).  In order to have sufficient time resolution (typically a tenth of a millisecond or less), it is necessary to zero-pad the frequency series prior to performing the FFT. Nonetheless, the computational cost is small.
Putting everything together we have
\begin{equation}\label{flike}
\ln L(\vec{\xi}, \vec{\theta})  = \sum_a 2 F_a \vert C_a(t_a) \vert \cos(\lambda_a - {\rm arg}\{C_a(t_a)\}) - \frac{F_a^2 D_a^2}{2 D^2_L}\, .
\end{equation}
The quantities $D_a$ and $C_a(t_a)$ can be pre-computed and stored for any choice of intrinsic parameters $\vec{\eta}$. The likelihood for any set of extrinsic parameters can then be found at the cost of a few multiplications and a cosine, allowing for millions of likelihood evaluations per second. Similar techniques can be used to accelerate the calculation of the extrinsic Fisher matrix, $\Gamma^E_{ij} = (\partial_{\xi^i} h | \partial_{\xi^j} h)$:
\begin{eqnarray}
&& \Gamma^E_{ij}  =  \sum_a  \left[ (F_{a,i}F_{a,j} + F_a^2 \lambda_{a,i}\lambda_{a,j}) H_{0a} \right.  \nonumber \\
&& \quad + \delta_{i t_c}(2\pi F_a^2 \lambda_{,i} t_{a,j}   \lambda_{,j} t_{a,i}) H_{1a}   \nonumber \\
&& \quad \left.  + \delta_{i t_c} \delta_{j t_c}(4\pi^2 F_a^2 t_{a,i} t_{a,j})H_{2a} \right]
\end{eqnarray}
where $H_{ka} = (f^k h_+| h_+ )_a$. The inner products $H_{ka}$ are computed once and stored. The luminosity distance can be extracted from the reference waveform by re-scaling the response function such that $F_a \rightarrow (D_a/D_L) F_a$, with $D_a$ scaled such that $H_{0a}=1$. 
The derivatives of arrival time at each detector, $t_{a,i}$, are non-vanishing for $\{\alpha,\beta,t_c\}$. The phase derivatives are non-vanishing for $\{ \alpha,\beta,\psi,\iota,\phi_c\}$:
\begin{equation}
\lambda_{a,i} = \delta_{i \phi_c} + \frac{ (F^a_{+}(\epsilon F^a_\times)_{,i}  - F^a_{+,i}(\epsilon F^a_\times) ) }{F_a^2} \, .
\end{equation}
while the derivatives of the re-scaled antenna pattern are non-vanishing for $\{ \alpha,\beta,\psi,\iota,D_L\}$:
\begin{equation}
F_{a,i} = -\frac{F_a}{D_L} \delta_{i D_L}+  \frac{ ( F^a_+ F^a_{+,i}  +  (\epsilon F^a_\times)(\epsilon F^a_\times)_{,i} )}{F_a} \, .
\end{equation}

\begin{figure}[htp]
\includegraphics[width=0.48\textwidth]{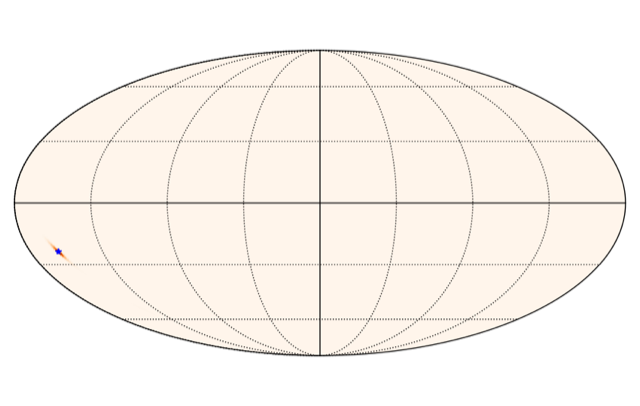} 
\caption{\label{fig:skyGW170817} Low latency sky map for GW170817. The blue star indicates the location of the electromagnetic counterpart to the BNS merger.}
\end{figure}

A PTMCMC algorithm is used to explore the extrinsic parameters. The initial ``burn-in'' phase can be accelerated by randomly trying out sky locations until one is found that yields the correct time delays between the detectors to within some pre-defined tolerance. However the likelihood evaluation is so fast that such acceleration is not necessary, and the chains can simply be initialized at some random draw from the prior distribution. During the burn-in phase the extrinsic PTMCMC uses the same number of chains and the same temperature ladder as the intrinsic PTMCMC from the coherent search. Each chain inherits the intrinsic parameters from the search phase. A mixture of proposal distributions are employed: Jumps along eigenvectors of the extrinsic Fisher matrix $\Gamma^E_{ij}$; small Gaussian jumps along each extrinsic parameter direction; and deterministic jumps along sky rings that preserve the time delay between a randomly selected pair of detectors~\cite{Cornish:2020dwh}. Samples from the chains with unit inverse temperature are used to produce low latency sky maps such as the example shown in Figure~\ref{fig:skyGW170817}.

\subsection{CBC Parameter Estimation}

The rapid coherent search and low latency sky mapping yield a good starting solution for a Bayesian exploration of source parameters. The inference is performed using the PTMCMC sampler Fisher matrix proposals, differential evolution, deterministic sky ring jumps in the extrinsic parameters and small Gaussian jumps along each parameter direction. The analysis is sped up by using a heteroydned likelihood function~\cite{Cornish:2010kf,Cornish:2020vtw}. 

The heteroydyned likelihood uses a reference waveform $\bar{h}$, in this case the maximum likelihood solution from the search, to re-write the log likelihood as
\begin{equation}
\ln L_a =   (\bar{r}_a | \bar{h}_a) +\frac{1}{2} ( \bar{h}_a | \bar{h}_a) - (\bar{r}_a | \Delta \bar{h}_a) - \frac{1}{2} ( \Delta \bar{h}_a | \Delta \bar{h}_a) .
\end{equation}
where $\bar{r}_a = d_a - \bar{h}_a$ and $\Delta h_a =   \bar{h}_a - h_a$. The $(\bar{r}_a | \bar{h}_a)$ and $( \bar{h}_a | \bar{h}_a)$ terms in the likelihood can be computed once and stored. The $( \Delta h_a|  \Delta h_a)$ term can be written as
\begin{equation}\label{hh}
4\int \frac{ \Delta h \Delta h^* }{S(f)} df =4 \int \frac{ \bar{{\cal A}}^2+ {\cal A}^2 - 2  \bar{{\cal A}}{\cal A}\cos\Delta \Phi}{S(f)} df 
\end{equation}
where we have used $h={\cal A}(f) e^{i\Phi(f)}$, $\bar{h}=\bar{\cal A}(f) e^{i\bar \Phi(f)}$ and $\Delta \Phi(f) = \bar{\Phi}(f) - \Phi(f)$. The phase difference between the reference waveform $\bar{h}$, and waveforms drawn from the posterior distribution $h$ will always be small, so using a reference waveform effectively heterodynes the numerator of equation (\ref{hh}), rending it a slowly varying function of frequency.  The $(\bar{r}_a |  \Delta h_a)$ in the likelihood can be written as
\begin{equation}\label{rh}
(\bar{r}_a |  \Delta h_a) = 4 \int   (\Re \bar{r}_w \Re \Delta h_w + \Im \bar{r}_w \Im \Delta h_w) df \, ,
\end{equation}
where
\begin{equation}
\bar{r}_w = \frac{\bar{r} \,e^{-i \bar{\Phi}(f)} S^{1/2}_s(f)}{S(f)}
\end{equation}
is the whitened reference residual heterodyned by the reference phase and
\begin{equation}
\Delta h_w =\frac{ \left(\bar{\cal A}(f)  - {\cal A}(f) e^{-i\Delta \Phi(f)} \right)}{S^{1/2}_s(f)}
\end{equation}
 is the heterodyned difference in the waveforms, whitened by the smooth component of the amplitude spectral density. The integrands in (\ref{hh}) and (\ref{rh}) can be written as products of a slowly varying function $s(f)$ and a rapidly varying function $r(f)$. In equation (\ref{hh}) the numerator is a slowly varying function, while the inverse of the full power spectral density is a rapidly varying function due to the spectral lines. In equation (\ref{rh}) the real and imaginary parts of $\Delta h_w$ are slowly varying while the real and imaginary parts of the heterodyned residual $\bar{r}_w$ are rapidly varying. The integrals (in practice sums over frequency) can be evaluated accurately and rapidly using a Legendre polynomial expansion. The sum over frequency is broken up into bands of width $\Delta f$ and the discrete Legendre polynomial expansions of the rapidly varying function $r(f)$ are computed once and stored for each frequency band. Each frequency band covers $M= T_{\rm obs} \Delta f$ frequencies, and the number of bands is $K= 2 f_{\rm ring}/ \Delta f$, where $f_{\rm ring}$ is the ringdown frequency of the reference waveform. 
The discrete values of the rapidly varying function in each band can be expanded in a sum of discrete Legendre polynomials:
\begin{equation}
r_k = \sum_{\ell=0}^M \rho_\ell P_\ell(k)
\end{equation}
where $P_\ell(k)$ are the discrete Legendre polynomials of order $\ell$~\cite{legendre} and the expansion coefficients are given by
\begin{equation}
 \rho_\ell = \alpha_\ell  \sum_{k=0}^{M} P_\ell(k) r_k \, ,
\end{equation}
where $\alpha_\ell$ is a normalization constant. The contribution to the inner products from each frequency band are given by
\begin{equation}\label{leg}
\sum_{k=0}^{M} s_k r_k \simeq \sum_{\ell=0}^1  \alpha_\ell^{-1} \rho_\ell \sigma_\ell \, ,
\end{equation}
where $\sigma_\ell$ are the expansion coefficients for the slowly varying function $s(f)$ and the sum has been restricted to just the first two terms in the Legendre expansion, which is usually sufficient when using short frequency bands, $\Delta f \leq 4$ Hz. For the slowly varying function $s(f)$ the required expansion coefficients in the $k < K$ frequency band are given by $\sigma_0 = (s((k+1)\Delta f) + s(k\Delta f))/2$ and $\sigma_1 = (s((k+1)\Delta f) - s(k\Delta f))/2$. The sum (\ref{leg}) includes the first and last bins in each frequency band, so there is a double counting of the contributions from these bins, which can be corrected for by subtracting the sum over the $K-2$ repeated values values. The heterodyning procedure speeds up the likelihood calculations by a factor of $\sim M$, with the largest speed up being for low mass, long duration signals such as those from binary neutron star mergers.

\subsection{Examples from GWTC-2}

To illustrate the performance of the sampler, two examples were chosen from the second Gravitational Wave Transient Catalog, GWTC-2~\cite{Abbott:2020niy}. The first example,
GW190924\_021846, was chosen as it was one of the signals flagged for glitch removal.The second example, GW190719\_215514, was chosen as it has among the lowest signal-to-noise ratios, and thus posed more of a challenge for the initial search. 

The prior ranges on the masses were set between $0.25 M_\odot$ and $150 M_\odot$. The priors on the aligned spin components were chosen to correspond to a uniform distribution of spin directions and magnitudes in an effort to mimic the priors used in the reference LIGO/Virgo analyses which using the IMRPhenomPv2 precessing spin model~\cite{Hannam_2014}. The analyses shown here used the IMRPhenomD phenomenological model~\cite{Khan:2015jqa}, which describes the dominant $\ell = |m| = 2$ mode of a quasi-circular, spin-aligned binary system. To stay within the domain of validity of this model the maximum spin magnitude was set to $\chi_{\rm max} = 0.85$ and the maximum mass ratio was set to $m_1/m_2 < 18$.

\begin{figure}[htp]
\includegraphics[width=0.48\textwidth]{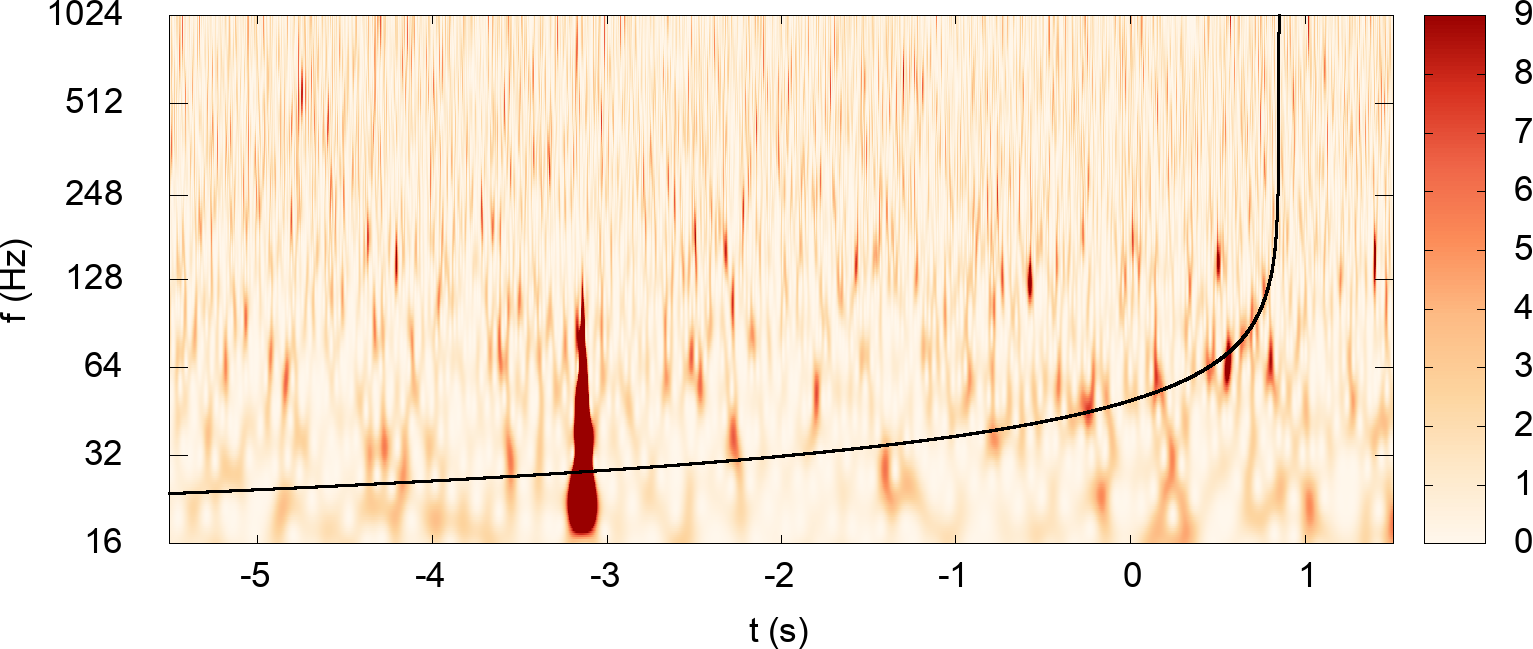} 
\includegraphics[width=0.48\textwidth]{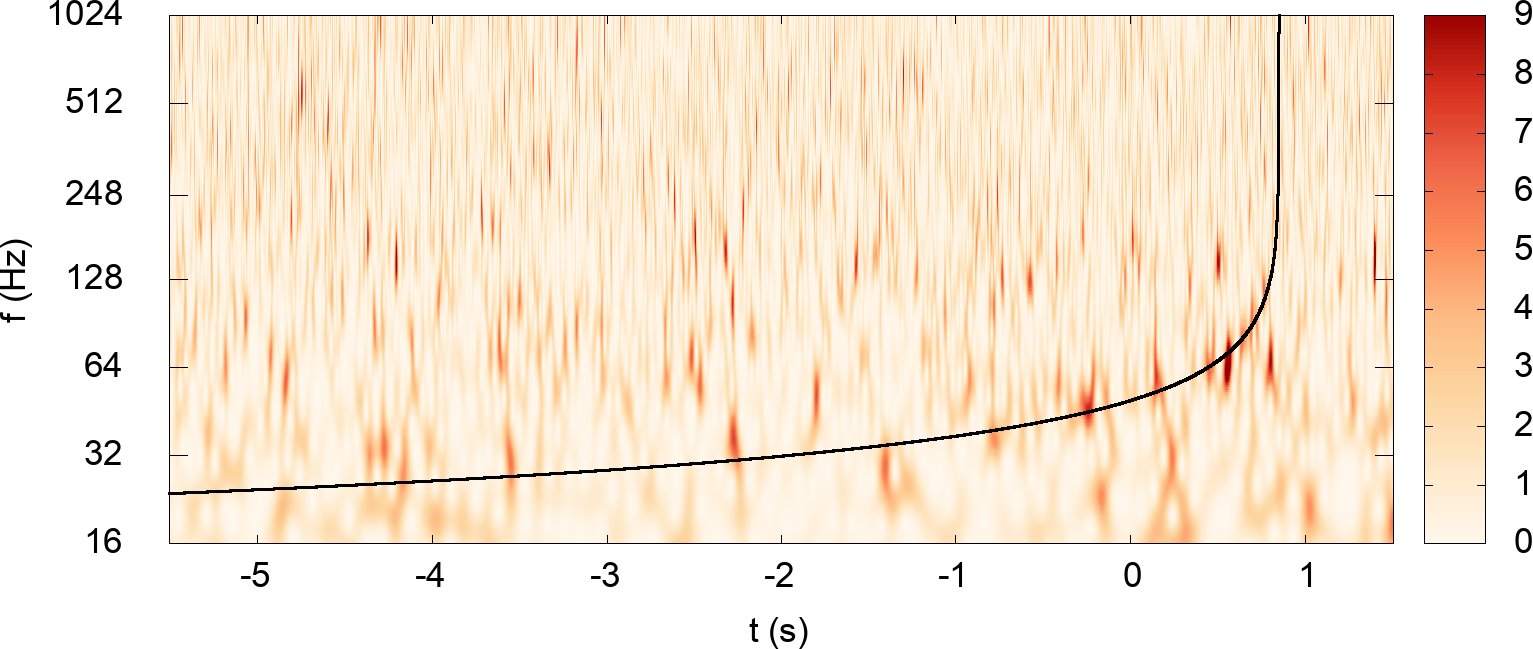} 
\caption{\label{fig:scanGW190924} Time-frequency maps of LIGO Livingston data centered at GPS time 1253326744. The black line indicates the reconstructed time-frequency track for GW190924\_021846. The upper panel shows the raw whitened data, while the lower panel shows the whitened data after wavelet de-noising.}
\end{figure}

Figure~\ref{fig:scanGW190924} illustrates the output of the search phase and glitch removal for GW190924\_021846. A moderately loud glitch that intersects the time-frequency track of the signal was identified and removed from the data. 

\begin{figure}[htp]
\includegraphics[width=0.48\textwidth]{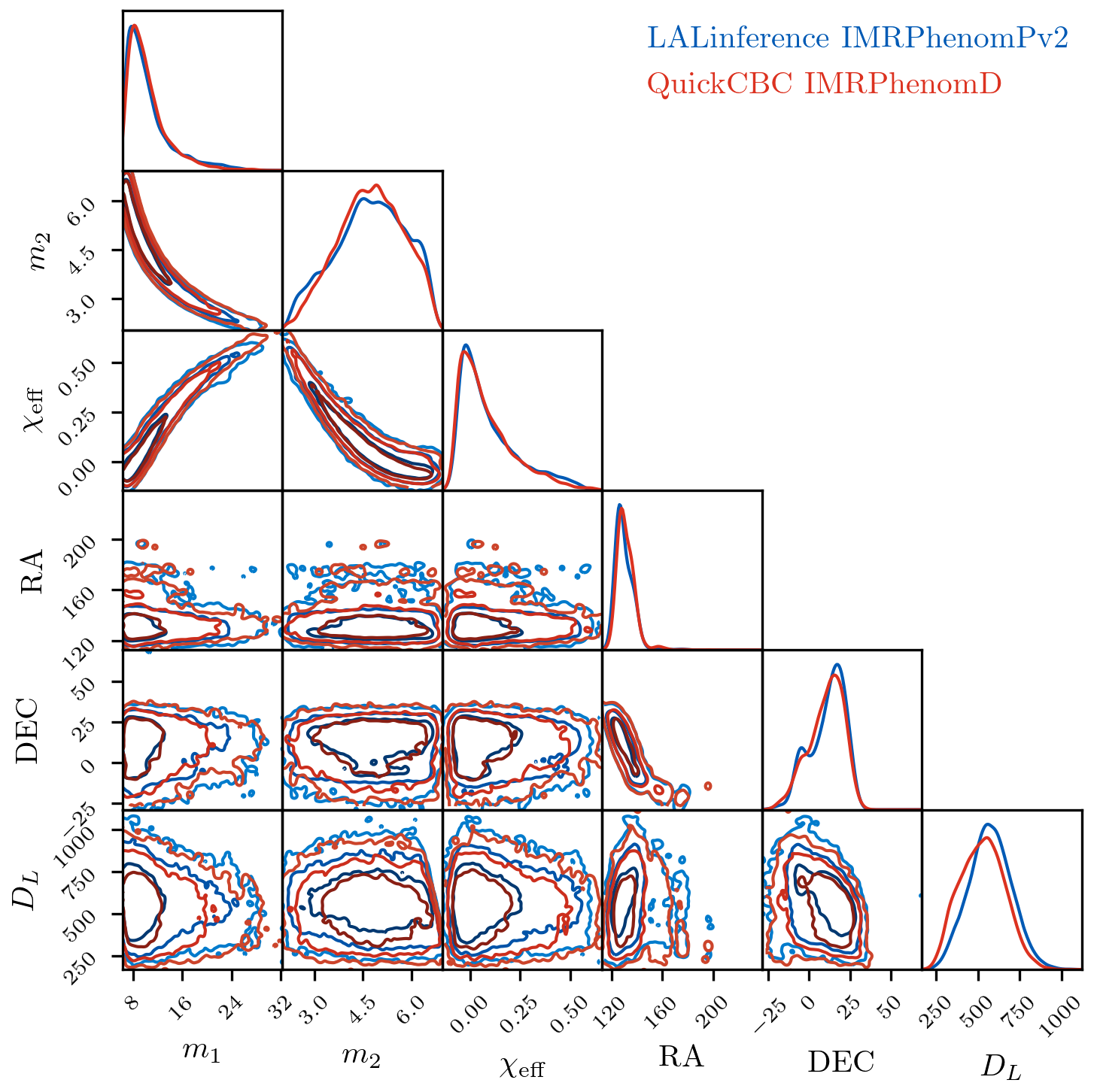} 
\caption{\label{fig:peGW190924} A comparison of parameter inference for GW190924\_021846 showing the preferred {\tt LALinference} IMRPhenomPv2 samples and the {\tt QuickCBC} IMRPhenomD samples.}
\end{figure}

Figure~\ref{fig:peGW190924} compares the output of the full {\tt QuickCBC} analysis using the internally de-noised with the publicly released {\tt LALinference} samples available from the GWOSC website. The {\tt LALinference} used a {\tt BayesWave} PSD and glitch model. The results show good agreement, the main difference being that the {\tt QuickCBC} analysis took a few minutes while the {\tt LALinference} analysis took a few days.

\begin{figure}[htp]
\includegraphics[width=0.48\textwidth]{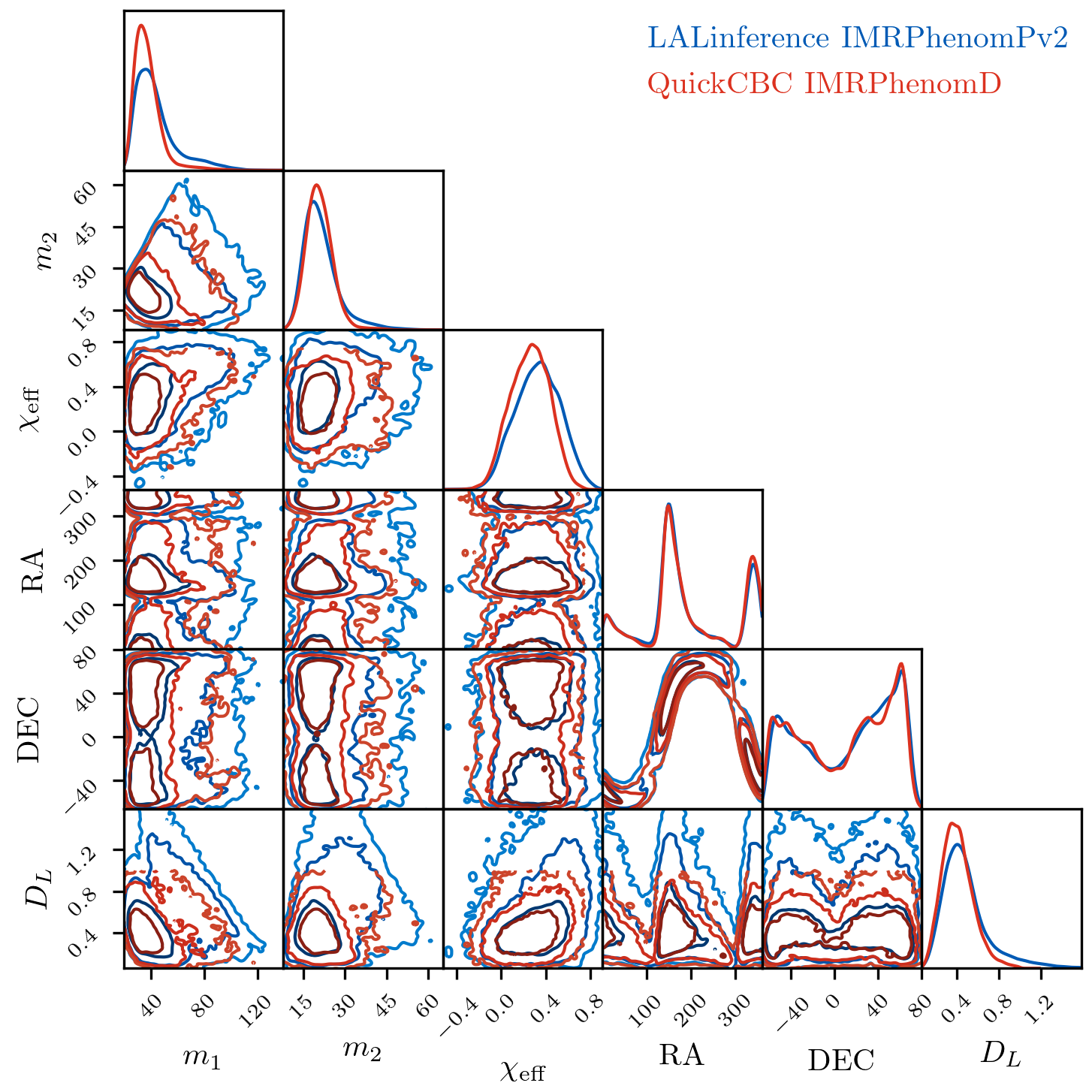} 
\caption{\label{fig:peGW190719} A comparison of parameter inference for GW190719\_215514 showing the preferred {\tt LALinference} IMRPhenomPv2 samples and the {\tt QuickCBC} IMRPhenomD samples.}
\end{figure}

Figure~\ref{fig:peGW190719} compares the {\tt QuickCBC} and {\tt LALinference} analyses for the low signal-to-noise ratio event GW190719\_215514. The weakness of the signal posed no obstacle to the {\tt QuickCBC} analysis, with the search phase locking onto the signal after a few hundred iterations. The posterior distributions from the two samplers are again in good agreement.

\section{Summary}

The {\tt QuickCBC} analysis pipeline is an end-to-end, open-source tool for gravitational wave data analysis. Its key features are speed and robustness against noise transients. The main limitation of the pipeline is that it currently only works with the IMRPhenomD waveform model. A near-term development goal is to expand the range of waveform models, starting with the IMRPhenomHM model~\cite{PhysRevLett.120.161102}, which includes contributions from higher modes, and the IMRPhenomD\_NRTidal model~\cite{Dietrich_2019} which includes tidal effects for binanry neutron star mergers. A longer term goal is to add precessing spin models.

Possible uses for the {\tt QuickCBC} pipeline for researchers outside the LIGO/Virgo collaboration are as a platform to develop novel analyses that can be applied to the publicly released data. Within the LIGO/Virgo collaboration the pipeline could be used to generate low latency sky maps, and to provide estimates for how likely it is that the system will result in the disruption of a neutron star, and thus a good candidate for producing an electromagnetic counterpart.

\section*{Acknowledgments}
The author is grateful for the support provided by NSF award PHY1912053. This work was initiated while the author was on sabbatical at the Observatoire de la C\^{o}te d'Azur, kindly hosted by Nelson Christensen. Discussions with Tyson Littenberg, Katerina Chatziioannou and Marcella Wijngaarden were very helpful. The author greatly appreciates Marcella Wijngaarden's help in tracking down an error in the sky localization algorithm, Charlie Hoy's help in extracting the {\tt LALinference} samples and Bence B\'{e}csy's for writing the scripts for automating the running of the pipeline in response to triggers from the Gravitational-Wave Candidate Event Database.  The author appreciates feedback on a draft version from Will Farr and Nelson Christensen. This research has made use of data obtained from the Gravitational Wave Open Science Center (https://www.gw-openscience.org), a service of LIGO Laboratory, the LIGO Scientific Collaboration and the Virgo Collaboration. LIGO is funded by the U.S. National Science Foundation. Virgo is funded by the French Centre National de Recherche Scientifique (CNRS), the Italian Istituto Nazionale della Fisica Nucleare (INFN) and the Dutch Nikhef, with contributions by Polish and Hungarian institutes.

\bibliography{refs}

\end{document}